\documentclass[format=manuscript, screen=true, review=false,nonacm=true, acmsmall, 10pt]{acmart}

\makeatletter
\newif\ifanonymous
\if@ACM@anonymous\anonymoustrue\else\anonymousfalse\fi
\makeatother

\usepackage{xcolor}
\usepackage{booktabs}
\usepackage{array}
\usepackage{hyperref}

\usepackage{csquotes}
\MakeOuterQuote{"}

\usepackage{enumitem}
\usepackage{pdflscape}

\usepackage{rotating}
\usepackage{changepage}

\usepackage{adjustbox}

\NewDocumentCommand{\rotNew}{O{45} O{1em} O{4cm} m}{\hspace{-#2}\makebox[#2][l]{\rotatebox[origin = br]{#1}{\begin{minipage}{#3}\hfill#4\end{minipage}}}}%

\title{Passing the Baton: Shift Handovers within Cybersecurity Incident Response Teams}

\author{Liberty Kent}
\affiliation{%
  \institution{University College London}
  \city{London}
  \country{United Kingdom}}
\email{libby.kent.23@ucl.ac.uk}

\author{Nilufer Tuptuk}
\affiliation{%
  \institution{University College London}
  \city{London}
  \country{United Kingdom}}
\email{n.tuptuk@ucl.ac.uk}

\author{Ingolf Becker}
\affiliation{%
  \institution{University College London}
  \city{London}
  \country{United Kingdom}}
\email{i.becker@ucl.ac.uk}

\begin{abstract}

Effective shift transitions are crucial for cybersecurity incident response teams, yet there is limited guidance on managing these handovers.
This exploratory study aimed to develop guidelines for such transitions through the analysis of existing literature and consultation with practitioners.
Two draft guidelines (A and B) were created based on existing literature and online resources.
Six participants from the UK and international incident response teams, with experience in shift handovers, were interviewed about handover structure, challenges, training practices, and their views on the draft guidelines. 
The collected data indicate the importance of signposting, evolving handover procedures, individual differences in handover style and detail, and streamlining the handover procedure. 
Participants agreed the drafts included all relevant details but suggested adding a post-incident review section and a service section for outages or technical difficulties. This study establishes a foundation for enhancing transition practices in cybersecurity incident response teams.

\end{abstract}

\begin{document}

% \received{XYZ}
% \received[revised]{XYZ}
% \received[accepted]{XYZ}

\maketitle
\newpage
\section{Introduction}
In the face of increasingly complex and sophisticated cyber-attacks, cybersecurity incident response teams (typically referred to as CSIRTs) are more important than ever. Incident responders face a cognitively challenging job; their tasks have been described as varied, often non-routine, and involve perception and comprehension of large amounts of information \cite{Erbacher_Frincke_Wong_Moody_Fink_2010,junior2023}. As such, passing the responsibility of these tasks from one worker to another during a shift handover (when workers finish their work shift and pass their current work to another member of the team) is particularly complex. The intricacy of incident response handovers is further exacerbated by incidents that span multiple days, necessitating the involvement of multiple teams who either work night shifts or operate under a ``Follow-The-Sun'' (FTS) model. The FTS model is a global workflow in which organisations pass an incident between multiple teams across multiple time zones so that teams only work during the daytime of their location~\cite{Fausett_Keebler_2022}. Communicating between multiple teams and potentially multiple time zones can present unique challenges~\cite{Fausett_Keebler_2022}, as handovers within-teams compared to between-teams may require different techniques. In CSIRTs, handovers may happen during cyber incidents, and could cover details on the current situation, what has been done already, steps currently being taken, and potential next steps. Despite the amount of research on handovers in other domains (see our literature review in Section~\ref{sec:literatureReview}), and their potential to create issue if done inefficiently, there is little research on shift handovers in CSIRTs. 

Drawing from the existing literature and online resources, we designed a set of guidelines for handovers in CSIRTs. The purpose of this initial exploratory study is to work with experts to improve on them. Due to the variety of roles and responsibilities within and across CSIRTs, the document is intended to be a template for those producing their own handover procedure for a team. Adjustment and adaptation will often be necessary in order to fit with a team's purpose, size, location, and other relevant factors. 

The rest of the paper is as follows: Section \ref{sec:literatureReview} of this paper will outline the literature relevant that formed the basis for this study, Section \ref{sec:Methods3} will explain the methods, Section \ref{sec:Results3} will cover the results, and Section \ref{sec:discussion3} will discuss the implications of the results, the limitations, and potential future directions.

\section{Related Works}
\label{sec:literatureReview}

\cite{Fausett_Keebler_2022} analysed handover processes in healthcare, nuclear power, railroads, and space shuttle mission control to identify factors that should be considered in handovers within SOCs. They state that the term SOC is used in the paper to describe ``a team of cybersecurity experts working together to defend an organisation from cyber threats'' (based on the definition by \cite{Zimmerman_2014}).

In healthcare, standardising handover procedure has been a priority for decades. It was issued as a national patient safety goal by the Joint Commission in the US in 2006~\cite{JointCommission_2012} and when investigated in 2012, it was found that 80\% of major errors could be attributed to miscommunication during handovers. \cite{Keebler_Lazzara_Patzer_Palmer_Plummer_Smith_Lew_Fouquet_Chan_Riss_2016} used meta-analysis to study the effects of these standardised handover protocols on patient, provider and organisational outcomes. The analysis demonstrated that these protocols improved results in all three outcomes, and the actual handover information passed. This applied across various levels of expertise, areas of clinical focus and provider type. They also found that more thorough protocols led to more information being passed, most significantly when including 12 or more items. They also found that training bundles in handover procedure improved outcomes significantly. 

Standardised handover procedures help develop shared mental models. This refers to an ``organised knowledge of structure of the relationships among the task the team is engaged in and how the team members will interact''~\cite{Salas_Rosen_Burke_Goodwin_2008}. It is widely accepted that shared team-based mental models relate positively to team performance (e.g.,~\cite{Mathieu_Heffner_Goodwin_Salas_Cannon-Bowers_2000}), meaning that we may see the improvement from standardisation in healthcare also occur in cybersecurity. \cite{Fausett_Keebler_2022} found that a checklist or handover toolkit may be an effective way to standardise procedure in SOCs. \cite{Steinke_Bolunmez_Fletcher_Wang_Tomassetti_Repchick_Zaccaro_Dalal_Tetrick_2015} also noted that many cyber-incident handling errors occur during handoffs. However, \cite{Nyre-Yu_Gutzwiller_Caldwell_2019} found that handovers within different cybersecurity organisations are highly varied in terms of procedure, formality, and documentation. Due to the variety of roles and responsibilities within and across CSIRTs and SOCs, the set of guidelines is intended to be a template for those producing their own handover procedure for a team. Adjustment and adaptation will often be necessary in order to fit with a team's purpose, size, location, and so on. This is common for checklists and programs within other domains (e.g., the I-PASS handover program is validated for physician handovers, but adaptations have been made to be implemented in nursing bedside reports, imaging/procedure handovers and other areas;~\cite{Blazin_Sitthi-Amorn_Hoffman_Burlison_2020}). 

\cite{Patterson_Roth_Woods_Chow_Gomes_2004} conducted observations across four high-consequence settings to identify handover strategies that could benefit healthcare systems. Their study of NASA Johnson Space Centre, two Canadian nuclear power plants, and railroad and ambulance dispatch centres yielded 21 distinct handover strategies (Table~\ref{tab:patterson-handover-strategies}). These strategies ranged from face-to-face verbal updates to delayed responsibility transfers during critical activities. While these settings share characteristics with CSIRTs, such as interconnected teams managing complex tasks, notable differences exist in terms of system infrastructure, digitisation levels, team sizes, and remote work capabilities. Additionally, CSIRTs may handle between-team handovers, particularly in Follow-The-Sun models, rather than solely within-team transitions. Therefore, this study examines which of Patterson et al.'s strategies are currently implemented in CSIRTs and investigates their practical effectiveness in modern cybersecurity operations.

\begin{table}[htb!]
    \centering
    \caption{Handover strategies across settings with high consequences for failure~\cite{Patterson_Roth_Woods_Chow_Gomes_2004}}
    \label{tab:patterson-handover-strategies}
    \small
\begin{tabular}{lp{11.6cm}} \toprule
No & Strategy                                                                                                    \\ \midrule
\multicolumn{2}{l}{Inferred objectives: Improve handoff update effectiveness}                                  \\ \midrule
1  & Face-to-face verbal update with interactive questioning                                                     \\
2  & Additional update from practitioners other than the one being replaced                                      \\
3  & Limit interruptions during update                                                                           \\
4  & Topics initiated by incoming as well as outgoing                                                            \\
5  & Limit initiation of operator actions during update                                                          \\
6  & Include outgoing team’s stance toward changes   to plans and contingency plans                              \\
7  & Readback to ensure that information was   accurately received                                               \\\midrule
\multicolumn{2}{l}{Inferred   objectives: Improve handoff update efficiency and effectiveness}                   \\\midrule
8  & Outgoing writes summary before handoff                                                                      \\
9  & Incoming assesses current status                                                                            \\
10 & Update information in the same order every time                                                             \\
11 & Incoming scans historical data before update                                                                \\
12 & Incoming reviews automatically captured changes to sensor-derived data before update                      \\
13 & Intermittent monitoring of system status while ‘on call’                                                  \\
14 & Outgoing has knowledge of previous shift activities                                                       \\\midrule
\multicolumn{2}{l}{Inferred objectives: Increase access to data}                                               \\\midrule
15 & Incoming receives primary access to the most up-to-date information                                         \\
16 & Incoming receives paperwork that includes handwritten annotations                                           \\\midrule
\multicolumn{2}{l}{Inferred objectives: Improve coordination with others}                                      \\\midrule
17 & Unambiguous transfer of responsibility                                                                      \\
18 & Make it clear to others at a glance which personnel are responsible for which duties at a particular time \\\midrule
\multicolumn{2}{l}{Inferred objectives: Enable error detection and recovery}                                     \\\midrule
19 & Overhear others’ updates                                                                                    \\
20 & Outgoing oversees incoming’s work following update                                                          \\\midrule
\multicolumn{2}{l}{Inferred objectives: Delay transfer of responsibility during critical   activities}           \\\midrule
21 & Delay the transfer of responsibility when concerned about   status/stability of process                   \\ \bottomrule
\end{tabular}
\end{table}

While Patterson et al.'s strategies remain relevant to modern cybersecurity operations, they require adaptation to account for technological advancement and industry-specific characteristics. Contemporary CSIRTs operate in highly digitised environments, often with distributed teams working remotely across multiple time zones. This study examines whether these strategies are implemented within current CSIRT operations to assess their practical effectiveness and identify necessary modifications.

While research specifically examining CSIRT handover procedures remains limited, relevant insights can be drawn from adjacent domains. National incident response protocols, such as those from the \cite{Australian_Signals_Directorate_Australian_Cyber_Security_Centre_2024}, outline communication requirements that parallel typical shift handover needs. Additionally, established handover practices from production and operations management~\cite{Beekeeper_2020} provide potentially transferable frameworks. These resources informed the development of our initial handover guidelines.

\section{Methods} 
\label{sec:Methods3}
This section describes the participant sampling strategy, materials developed for the study, and procedural approach. The study was approved by University College London Crime Science ethics committee.

\subsection{Participants}
The study comprised six participants. Participants were recruited through adverts posted on professional social media, with the inclusion criterion of current CSIRT employment and experience with shift handovers. Participation in the study was voluntary with no financial incentive offered.

\subsection{Materials}
\label{sec:materials}
Two drafts of handover materials were developed as a prompt to provide a starting point for participants to build on. Guideline A was created using the Australian Signals Directorate cybersecurity incident triage questions and situation report template \cite{Australian_Signals_Directorate_Australian_Cyber_Security_Centre_2024}. The triage questions are specifically for when an organisation is reporting an incident to the Australian Signals Directorate, and the situation report template gives no guidance as to which situation this template should be used. However, the questions appear to be applicable to a shift handover. This formed the first guideline (Table~\ref{tab:guidelineA}). It was formatted as a checklist of questions that an outgoing member of staff would simply indicate as to whether their handover had covered it, and any extra comments they had. This meant that the checklist was not intended to serve as a method of communication in itself, and was to be used alongside either a verbal or written handover to verify that all relevant information and materials had been passed on.

\begin{table}[htbp!]
    \centering
    \caption{Guideline A: Column 1. (The full handover guidelines would contain two additional columns, one to tick if the item has been discussed, and one to enter additional comments if necessary.)} 
    \label{tab:guidelineA}
    \small
\begin{tabular}{  m{13.5cm}  } \toprule
Item
\\ \midrule
What is the event?
\\
When did the event occur?
\\
When was the incident discovered?
\\
How was the incident discovered?
\\
What assets are being affected?
\\
What vectors of attack are associated?
\\
What is believed to be the source/cause of the incident? 
\\
What are the current impacts?
\\
What are potential impacts, and which present greatest concern?
\\
What are our priorities?
\\
What is the incident severity?
\\
What has been done so far, when, how, why, and by whom?
\\
What data and information are you handing over?
\\
What is the current plan for next steps, and timeline?
\\
Are there any inter-team collaborations (external parties, other departments), what do they know, what has been done, what may need to be communicated further?
\\ 
Has law enforcement been contacted, what has been done, what do they know, what may need to be communicated further?
\\
Who are the current and previous call attendees, and their roles, responsibilities and chain of command?
\\
Who are the incoming call attendees, and their roles, responsibilities and chain of command?
\\
Is there anything else relevant or necessary to mention?
\\
What other comments do you have on the situation?
\\
What are the contact details for team members and any other collaborators?
\\
What is the incident ID?
\\ \bottomrule
\end{tabular}
\end{table}

Following the development of Guideline A, we created Guideline B with a more comprehensive structure based on shift handover templates from production plant operations.

This second guideline (Table~\ref{tab:guidelineB}) was more detailed and formed more of a handover template than checklist, in order to closer replicate the example handovers such as ~\cite{Beekeeper_2020}.
This meant that it was not necessary to also use another means of communication, verbal or written: the guideline served as a way to convey information, rather than simply verifying that information had been conveyed in some manner. Both guidelines were presented to participants in order to gather preliminary information on which layout was optimal, as there was little information available as to what this material should look like.

\begin{table}[htbp!]
    \centering
    \caption{Guideline B}
    \label{tab:guidelineB}
    \begin{adjustbox}{max width=.9\textwidth}
\begin{tabular}{  lm{9.8cm} } \toprule
Item & Description                                                                                              \\ \midrule
\textbf{1. General Information} & 
\\
Shift Date and Time & 
\\
Outgoing Team Lead & 
\\
Incoming Team Lead & 
\\
Shift Duration & 
\\
\textbf{2. Current Incident Status} & 
\\
Active Incidents & 
\\
Incident ID & 
\\
Description & [brief description]
\\
Severity Level & [Low/Medium/High/Critical]
\\
Affected Systems & [List of systems]
\\
Current Status & [Investigation, Mitigation, Remediation, etc.]
\\
Assigned Team Members & 
\\
\textbf{3. Actions Taken During Shift} & 
\\
Detection & [List all detections during the shift]
\\
Analysis & [Summary of analysis performed]
\\
Containment & [Actions taken to contain the incident]
\\
Eradication & [Steps taken to eradicate the threat]
\\
Recovery & [Recovery actions and status]
\\
Communication & [Stakeholders informed and communication summary]
\\
\textbf{4. Pending Actions and Follow Ups} & 
\\
Pending Analysis & [List items requiring further analysis]
\\
Pending Containment & [List containment actions pending]
\\
Pending Eradication & [List eradication tasks pending]
\\
Pending Recovery & [List recovery steps pending]
\\
Pending Communication & [List stakeholders who need updates]
\\
\textbf{5. Escalations and Key Decisions} & 
\\
Escalations Made & [Details of any escalations]
\\
Decisions Taken & [Summary of key decisions and their rationale]
\\
\textbf{6. Documentation and Reporting} & 
\\
Incident Reports & [Status of incident reports]
\\
Documentation Updated & [Any updates to documentation or playbooks]
\\
Logs and Evidence Collected & [Summary of logs/evidence collected]
\\
\textbf{7. Briefing and Handover Meeting} & 
\\
Handover Meeting Time & 
\\
Participants & [Names of outgoing and incoming team members]
\\
Summary Discussion Points & [Key points discussed during the handover meeting]
\\
\textbf{8. Additional Notes and Observations} & 
\\
Outgoing Team Notes & [Any additional notes from the outgoing team]
\\
Incoming Team Acknowledgment & [Confirmation from the incoming team that they have received and understood the handover]
\\
\textbf{9. Contact Information} & 
\\
Emergency Contacts & 
\\
Key Stakeholders and Contacts & 
\\
\textbf{10. Sign Off} & 
\\
Outgoing Team Lead Signature & 
\\
Incoming Team Lead Signature & 
\\
Date and Time & 
\\ \bottomrule
\end{tabular}
\end{adjustbox}
\end{table}

The development of two distinct formats enabled us to evaluate industry preferences between a supplementary verification tool and a comprehensive handover template. 
While Guideline B could supplement other communication methods, it was designed to function independently, unlike the verification-focused Guideline A. Previous research had indicated that more details were beneficial, but had also indicated a lot of variances between teams so it was necessary to use both to identify a balanced point between enough detail without being too specific to use across different team purposes, functions and sizes. 

To evaluate these guidelines, we developed a structured interview protocol (Table~\ref{tab:questionlist}).

\begin{table}[htbp!]
    \centering
    \caption{Interview Question List}
    \label{tab:questionlist}
    \small
\begin{tabular}{ m{13.5cm} } \toprule
Give a brief explanation of what your organisation does.
\\
Give a brief explanation of what your role includes.
\\
How much collaboration is involved?
\\
How would you classify the size of your organisation? (less than 10, 10-50, 50-100, 100-1500, 1500+)
\\
Have you dealt with a large cybersecurity incident within the last 5 years? What was the impact of that incident?
\\
How did you deal with it?
\\
Does your role involve shift handovers?
If not, why not?
\\
What should a shift handover include? 
\\
What is the main goal of a shift handover?
\\
What do your current shift handovers look like (if applicable)? How are they documented?
\\
Are they individual or group-wide?
\\
Are they verbal or written?
\\
What would an optimal shift handover look like to you?
\\
How do shift handovers fail?
\\
What difficulties are faced when completing a shift handover?
\\
What could help overcome these difficulties?
\\
Which of these sets of guidelines look the closest to something that could be useful to those writing their own shift handover procedures?
\\
Which elements of these drafts should remain included in the next draft (i.e., which elements are useful)? 
\\
Which elements need to be adjusted or removed?
\\
Are there any missing elements from the drafts?
\\
Are there any other improvements to be made that haven’t been covered?
                  \\ \bottomrule
\end{tabular}
\end{table}

\subsection{Procedure}

Participants received an information sheet and consent form prior to the study. Following consent, we provided the interview questions in advance to allow participants to familiarise themselves with the topics and, particularly for those under non-disclosure agreements (NDAs), prepare responses that adhered to their confidentiality requirements. Participants could decline to answer any question (although none did so). The semi-structured interview format was explained to participants at this stage.

Each participant completed a one-hour semi-structured interview conducted via Microsoft Teams. The interviews followed the prepared questions while allowing for relevant follow-up inquiries. As highlighted in Table~\ref{tab:questionlist}, topics included background information about their job and experience, what handovers look like for them, how they feel about the process, what difficulties arise, how to address these, and feedback on the guidelines. Interviews were transcribed through Microsoft Teams.

We analysed the data using an inductive thematic approach~\cite{Thomas_2006}. This consists of familiarising with the data, searching, reviewing and defining themes, and creating categories~\cite{Braun_Clarke_2006}. All authors carefully re-read the transcripts. The first author then coded the data using Microsoft Excel by picking out quotes from the transcripts that were relevant and captured responses to each question, or unique points from a response if there were more than one. 
These codes were initially grouped by interview question and subsequently analysed to identify recurring themes across participants' responses. The grouping and themes occurred in an iterative process, regularly discussing all decisions between all authors and reviewing the quotations throughout.
We do not report inter-rater agreement scores, as they are inappropriate in TA~\cite{clarkebraun2021}.

\section{Results}
\label{sec:Results3}

The participants represented diverse geographical locations, including the United Kingdom (N=3), Morocco (N=1), the Netherlands (N=1), and Saudi Arabia (N=1). Four participants were involved in all aspects of incident response, while two specialised in detection and analysis. Five participants worked in organisations providing cybersecurity as a service, with one representing an internal CSIRT in the public sector.

We thematically analysed the six transcripts and identified 81 relevant quotes. We systematically coded and analysed the quotes, resulting in the emergence of 11 distinct themes. These themes were then mapped to our interview questions from Table~\ref{tab:questionlist}, as shown in Table~\ref{tab:results}.

\begin{table}[htbp!]
    \centering
    \caption{Table of Questions, Quotes and Themes}
    \label{tab:results}
    \small
\begin{tabular}{ m{5cm}m{1.5cm}m{6cm} } \toprule
Question & Number of Quotes & Themes identified within answers \\ \midrule
What does a handover include? & 7 & Signposting
\\
Is this optimal handover? & 5 & Evolution
\\
Is what to include taught, or do you rely upon intuition? & 5 & Informal teaching
\\
Is there much variation? & 5 & Individual differences
\\
What is the main goal of a shift handover? & 4 & 
\\
Difficulties faced? & 16 & Missing and outdated information, longer incidents, remote work, technical issues
\\
What could overcome these? & 6 & 
\\
Which guideline looks best? & 6 & 
\\
Which elements of draft are useful? & 13 & Conveying rationale, different organisational contexts
\\
Which elements of draft should be removed/adjusted? & 14 & Streamlining
\\ \bottomrule
\end{tabular}
\end{table}

\subsection{What does a handover include?}
Participants had varying answers about what a handover included in their team. One participant explained their system, that composed of 
\textit{``three different [sections]. So security issues, service issues and any miscellaneous sort of notes.''}

Another spoke of their view of handover contents, saying that they should cover 
\textit{``what assets are compromised? Accounts that are compromised? The main communication with the clients, the story of the client. […] Maybe a documentation of the evidence gathered.''}

The other four responses highlighted a common theme. One responded 
\textit{``[technical details] would be in a ticket. So quite often when we hand over, we put loads of ticket references in so that way it keeps everything in one location and you can add comments on top with a timestamp to see who's worked on it and at what times. [Handovers are] more for signposting.''} The other response was 
\textit{``[you have] 30 minutes to document everything, to talk about what you have done, and then anything for ongoing investigation. The handover also includes passing the tickets to the next team that are going to manage that.''} 
These indicate a theme of \textbf{signposting}, where a ticketing system may be employed to work alongside the handover process to reduce the amount of information in the handover alone.
This ticketing system can allow multiple members of a team to view and edit the same bank of information about an incident, regardless of who handed over to them, which aligns with Patterson et al.'s (\cite{Patterson_Roth_Woods_Chow_Gomes_2004}) strategy 19.

\subsection{Is this handover process optimal?}
Most participants considered their handover processes to be reasonably optimal, with several noting ongoing \textbf{evolution} in their procedures. As one participant explained:
\textit{``I think so, in that we matured over quite a bit of time with improvements and better e-management tools.''} 
However, one participant from an organisation with infrequent handovers offered a contrasting perspective:
\textit{``because it doesn't happen that often, I don't think we paid a lot of attention in developing a proper handover. So I wouldn't say it's optimal.''} 
This organisation's structure has different teams for different stages of incident response, and information is passed between them as responsibility crosses over from one team to the next, but this was explained to be viewed as a very different process to when large incidents necessitate shift handovers within one team. The process of handing responsibility over in the manner this paper discusses occurs rarely enough in this company (approximately once every three months) that they have not developed handover procedure. The reportedly very different process of information sharing between the company teams was not explored. 

\subsection{Is what to include in handover taught, or do you rely upon intuition?}
Of the four participants who addressed this question, two held training responsibilities within their organisations so could provide more insight as to the thinking behind the decision. These participants reported that new analysts had no formal training specifically on handover procedure, but did have some degree of \textbf{informal training}; this included shadowing seniors while they produced their handovers, and seniors reading and feeding back on the new analysts' own handover notes. One participant explained 
\textit{``during the day, I will highlight [things that should be included] myself to put in the handover. So you can start to learn through me telling you what I want covered. There's a big difference between me telling them to do it and then doing it, and them going off their own back and writing stuff down.''}
This approach of learning through example was also seen by the other participants, who said 
\textit{``I do incorporate that into the training. And I'm not saying that you should write an example handover and I'll check it. The methods I get across to them is `if you were on the receiving end, what information would you like? Given what I've taught you?'''} 

\subsection{Is there much variation in handovers between individuals?}
The informal training approach appeared to contribute to significant variation in handover practices between individuals. Participants identified several factors influencing these \textbf{individual differences}, particularly focusing on experience levels and staff turnover.
One participant specifically spoke about the experience levels of team members affecting their intuition over what to include, saying 
\textit{``Guess it is a judgement call, right? If you've got a newer guy and he's writing the handover, maybe it's not [intuition]. A couple of new guys... in the handover, they'll talk about anything and everything that's happened. [And then] you've got the exact opposite [where they say] something did happen but nothing really interesting [came of it], so they're not going to talk about it.''}
Another participant discussed how the high turnover rate in cybersecurity affects handover quality: 
\textit{``it depends on the person. Because when we have someone motivated and interested by the job and everything, they do it as requested and give more [detail]. But sometimes we have these turnover, so sometimes the people are not interested, they only [include what they] feel is the minimum requirement.''}
In contrast, one organisation reported more standardised practices:
\textit{``It's fairly standard, everybody does mostly the same. [And for the tickets] we have really, really tight standard operating procedures and for each ticket we have a lot of automation as well.''}
This uniformity was attributed to limiting handover responsibilities to senior staff members with extensive handover experience.

\subsection{What is the main goal of a shift handover, and what does this look like practically?}
Experience of handover procedure between participants varied, potentially due to team size, location, and average severity of incidents seen. One participant explained that handovers are not often seen outside of more severe and time-critical incidents, as it is easier to have the same group of analysts complete an analysis themselves. As such, their procedure is simple and concise:
\textit{``Time is not an option you have here, but you also know your [team members] have to take some time off. It's not often we see this in our incidents. When we need it, when [an incoming team] come in the office, the first team [is] there. So they describe to us what they think, where they want us to go.''}

Face-to-face handovers were particularly emphasised by participants working on-site with clients:
\textit{``it's very important, in order to share more details and information across teams. We are working on site, so each time we meet at the end of the shift, it starts for [the next team].''} These face-to-face meetings align with Patterson et al.'s (\cite{Patterson_Roth_Woods_Chow_Gomes_2004}) first strategy for effective handovers.

In contrast, other participants reported relying primarily on written communication. One participant emphasised information accessibility: \textit{``you can publish it all from Teams and read all the previous handovers if you want to as well.''} While this participant noted some challenges with ensuring accurate information transfer through written communication, the approach enabled implementation of Patterson et al.'s (\cite{Patterson_Roth_Woods_Chow_Gomes_2004}) strategy 19, allowing team members to access historical handover information from multiple sources.
The core goal of handovers remained consistent across approaches, as summarised by one participant: \textit{``it's making sure that everyone's on the same page. It's important that the night shift understand what's going on, if anything needs to happen. And what the next steps are for us to be doing. Giving a bit of forewarning.''}

\subsection{What difficulties are faced?}
This group consisted of the most quotes, giving substantial insight into the complexity of designing an optimal handover process. 4 key themes arose here, which are explained below.

\subsubsection{Missing and Outdated Information}
Several participants noted the concern of missing information during handovers. One spoke of when others may give incomplete handovers, saying 
\textit{``it's still reliant on the person writing the handover. In a way, it's not foolproof: if they don't put down what they should be putting down, then the next team isn't aware, right?''} Information can be forgotten or missed in many ways; whether this is massive cognitive load leading to missed detail, fatigue at the end of a shift leading to carelessness, or even just poor communication skills. 
Another spoke of this worry while writing a handover oneself, explaining 
\textit{``sometimes you'll forget to add a part of the investigation that you're some notes you've got, or maybe attach some like log files you've got. Yeah, it's fast-paced.''}

This concern was also reflected in a discussion of keeping up with the fast pace mentioned. Another participant explained just how quickly information can change, and how constantly updating notes over long shifts can be difficult:
\textit{``going to update the handover, let's say at 9am you've got something to put in the handover. By 9:30, that might have changed. Because the analysts are doing 12 hour shifts, to remember everything that's happened during the day and add it in, [there's] bound to be something you'll forget.''}
This was similarly talked about by another participant, who discussed how the workload can prevent handover notes from being updated as often as they wished, especially when exacerbated by busy shifts and limited staff. 
\textit{``During the shift there can be too many incidents, too many requests, too [few] staff… it's sometimes hard for the teams to [write notes]. So they don't necessarily fill it in right away.''}

Fatigue and time pressure also impacted handover quality. One participant observed: 
\textit{``over-utilised analysts are just gonna be ready to just get out and head home. So they just wanna get it done fast, and rush.''}

Not only this, but incoming staff have limited time to absorb and process all the information in a handover before the outgoing staff leave, meaning they have to do this very quickly in order to notice any gaps in their understanding and ask questions to clarify while they have the chance. One participant noted 
\textit{``[I wish] the teams were briefed earlier on that we will be taking over, they would be prepared [...and know] to ask the right questions and what they need to know.''}

\subsubsection{Longer Incidents}
Incidents spanning multiple shifts presented distinct challenges for handover effectiveness. Many incidents can span multiple weeks or months; for example, in 2024, the average downtime after a ransomware attack was 3.4 weeks ~\cite{Veeam2024}. A key concern was the accumulation of handover reports and the difficulty of maintaining context across extended incidents. Often, incoming teams must read multiple handover reports in order to be fully briefed on a longer incident:
\textit{``say there's been [an incident across] three days and three nights. Unless you go back and read, what, 6 handovers? You don't actually have that context of the week.''}

This challenge of maintaining historical context was echoed by another participant who highlighted how information engagement tends to decrease over time:
\textit{``The first day, second day, third day, they will start to ignore [the older handovers]. The continuous nature means you get more details about what happened and what [the collaborators] are working on, their motivation and everything... you [are] adding more and more information.''}
This observation suggests that while longer incidents generate more detailed documentation, teams may struggle to effectively utilise this growing body of information during handovers.

\subsubsection{Remote Work} 
Remote operations introduced additional complexity to handover processes, particularly regarding communication effectiveness. Participants working remotely explained how written communication specifically was challenging, especially when face-to-face interaction is not possible:
\textit{``[there is] worry of making it clear enough that it can be interpreted by everyone, compared to doing it in person. Sometimes to get that across over text is quite hard and you only got such a limited amount of space.''}
When asked why they were not communicating verbally, they clarified that many worked from home, and verbal conversations over the phone would be disruptive due to the time of day that they would often be carried out.
\textit{``I'd prefer doing this stuff in-person [or at least verbally] but that sort of aspect is ruled out. Working from home and you're doing a 7am handover, whispering isn't great. If you've got people at home that are still asleep. [And] it would be unlikely that we'd ever move back to it in person.''} This suggests that remote work has created a persistent challenge requiring teams to optimise written communication methods rather than rely on traditional verbal handovers.

\subsubsection{Technical Issues}
Participants identified two distinct technical challenges that impact handover effectiveness: access management and software limitations for documentation.

Access management issues could disrupt handover execution when assumptions about capabilities did not align with reality. One participant explained:
\textit{``[another] thing that can go wrong is if we hand over, say, a request that we believe the analysts can do, but it turns out it's out of our remit. Whether that's just because it's not our job, or lack of permissions and lack of access to resources.''}

Another participant explained the importance of using sufficient software to support handovers. Not only do technical details need to be covered, but analysts need to have facilities to explain their reasoning and add extra details: 
\textit{``if your technology does not support the tracking of your analysis in terms of tagging, in terms of noting, it makes it harder for the second team to know what the first team have reached in terms of why they concluded that.''} This highlights how technical infrastructure must support not only the documentation of actions taken, but also the reasoning behind analytical decisions.

\subsection{What can be done to overcome these?}
Despite the number of difficulties identified, some solutions were suggested to improve handovers. Two of these were techniques already implemented by some participants that addressed difficulties brought up by other participants, and two were hypothetical but seemed plausibly practical.

One participant, responsible for their team's handover procedure, explained that he adjusted shift cycles in order to have more analysts in on busier periods, and more shift overlap between incoming and outgoing teams:
\textit{``What I've started to do is I've pulled resource off nights, I've pulled resource off the weekends, and I've sort of put them on cycles of [varying shifts]. So you kind of got most resources in the afternoons where we're busiest, and then it sort of drops off.''}

This helps alleviate the negative effects of overwork from shifts being understaffed, as the shift pattern allows times of peak demand to be covered by more team members and the work distributed between them. Additionally, this process allows some incoming individuals to work alongside the outgoing members for a while before they leave, meaning this addresses concerns over having ample opportunity to ask questions and clarify information.

Another implemented solution focused on streamlining the process of handover note-taking by allowing the notes over the day to actually form the handover itself, instead of needing to write something up at the end of the shift:
\textit{``I've got it set up so it integrates with Teams, so you can actually write it in Teams as the day's going on.''}
This helps with the concern of keeping up with the fast pace, as the more convenient and quicker it is to take notes during a shift, the easier it is to take continual notes through the day that cover all the progress made.

One proposed solution was also proposed by a participant in charge of their team's handover procedure, who said:
\textit{``[we could use] a mechanism where you could put something in [a handover] that would stay until someone got rid of it. It's that mechanism of rather than a 12-hour summary, it's more what's happened over the last set of shifts since I've been in.''}
This referred to the prospect of creating the option for the last handover to also appear in the next handover when an ongoing incident is happening, and to keep collating previous handovers within the current handover until the ongoing incident is resolved and the mechanism is turned off. This addresses the challenge mentioned of incoming team members having to maintain context by reading previous handovers outside the one they are being actively involved in at the time, even if an active incident has spanned other handovers that therefore include details potentially relevant to their work. 

The other proposed solution focused on improving handover preparation: 
\textit{``[incoming individuals] knowing earlier about the investigation and about the findings and being up to date might help the team to ask the right question in case of the first team who was over-utilised and tired to ask the right question.''}
This may not be feasible for some companies as those not currently on shift may not want to be required to read a ticket, handover or report outside of their paid hours, and the company may not be able to afford to pay for any extra time. However, the approach aligns with Patterson et al.'s strategy 11 (and possibly 13) (\cite{Patterson_Roth_Woods_Chow_Gomes_2004}) suggesting its viability in some organisational contexts, so it should be viable for some CSIRTs too.

\subsection{Which guideline looks best?}
This question received an almost unanimous answer of ``B''. As such, the following feedback will be (unless otherwise stated) regarding Guideline B. 

Several participants offered contrasting views about organisational fit (i.e., internal and external SOCs, MSSPs, internal and external CSIRTs). One participant noted the potential for Guideline A being utilised by SOCs, saying 
\textit{``Guideline A, it didn't feel relevant to me as an incident responder. However, it is very relevant to the SOC.''} 
However, another participant who worked in an external SOC that only offered detection and analysis (i.e., not incident response) said both guidelines looked too structured for their liking:
\textit{``So I guess, in incident response, you are focused on responding to an attack, even may have multiple areas covered, [so this may work for them] but this stuff is obviously far more specific than is necessary for me.''} 
This participant went on to say 
\textit{``[Guideline B] would be maybe more a handover I would expect for an internal SOC rather than an MSSP. I think this stuff is probably super key if you've got a 24 by 7 internal SOC where you're fully managing all of this stuff yourself.''}
In contrast, another participant who worked in an MSSP explained 
\textit{``guidelines B---response team shift handover checklist---it's something similar that we are doing.''}
This participant also later said 
\textit{``[A could be used for measuring] growth, how things are going, so they can look at it metrics-wise. So it could be that both of them could be used to complement each other, but geared for different audiences.''}

Notably, the sole participant from an internal CSIRT preferred Guideline A, explaining:
\textit{``I would choose a Guideline A instead of B, because with B, most of the information is going to be in the in the ticketing system already.''}

\subsection{Which elements of the draft are useful?}
Participants identified several valuable elements in the guidelines, with a particular emphasis on sections that facilitated the communication of analytical reasoning and decision-making processes.
The importance of \textbf{conveying rationale} emerged as a key theme. One participant highlighted the value of the detection section for explaining analysis decisions:
\textit{``[The] detection [section] during the shifts is good so that the incident responder will say this is why we sent it to the client, why we didn't send it, why this is false positive, why this is a true incident and everything.''} 
Similarly, another participant discussed section 4's role in communicating tactical decisions: 
\textit{``I really like Section 4 with the follow-ups because... you do have to explain, you know, [where someone is coming in]. If we're doing handover during containment we go, okay, these particular boxes have been isolated. At this point, now the focus is to make sure that whatever it was that ransomware and it's not been able to spread and we need to focus on that and then look at eradication.''}
This need to convey not only objective technical details but also opinion and rationale aligns with strategy 6 of Patterson et al.'s (\cite{Patterson_Roth_Woods_Chow_Gomes_2004}).

Participants also identified specific sections as particularly valuable for \textbf{different organisational contexts}. 
For MSSPs, Sections 1 and 2 were considered essential: 
\textit{``the first section for me is something very important. Especially for MSSPs because sometimes you don't have visibility on who works on what.''}
\textit{"so Section 2 is a given. That's 100\%, we'll need that because that's the point of the handover, right."}
On top of this, one participant explained their opinion on the structure of Guideline B:
\textit{"the one worry that I've had [in the past] is the tick box element... quite a lot of people just take it as almost like a chore […] when you've got that element of description, it encourages you to add a comment to it. It's just got that detail that encourages you to enrich it and to provide value when handing over."}

\subsection{What needs to be adjusted?}
This question once again led to many relevant points, but only one common theme: \textbf{streamlining}. Participants gave suggestions on how to condense the guidelines in order to be more efficient, saying 
\textit{"so something like this it would take a long time to write... priority is making it as streamlined as possible, make it quick and easy, make it cover all the relevant bits and get rid of all the fluff."}
This often led to discussion on which elements could be removed specifically in the case of their own organisation. However, one recommendation was generalisable: 
\textit{``section two and three, you could probably blend them together- make it a little bit bigger in the comments and say the action taken.''}

Two participants proposed additional sections to enhance the guidelines' utility. One suggested including operational status information:
\textit{``I guess, I think my only thing [missing] would just be the sort of service side of stuff. So if we've got any outages?''}
The other suggested addition was a post incident review section, including 
\textit{``any lessons learned from the incident when you close it down.''} 
These suggestions indicate potential gaps in the current guidelines' coverage of operational status and knowledge management.

\section{Discussion}
\label{sec:discussion3}
From these results, we can see that the majority of participants consider all relevant details to be covered by the guidelines, with the remainder providing two suggestions for additional sections (a service section and a post incident review). Clear preferences emerged regarding guideline format and application.

\subsection{Which guideline is best?} 

As discussed in Section~\ref{sec:materials}, Guideline~\ref{tab:guidelineA} and Guideline~\ref{tab:guidelineB} vary in their structure, content and intended usage; Guideline B serves as a means of communication in itself, by filling in the template with the prompted information and passing it on (although this can be used in conjunction with other means) while Guideline A serves as simply a checklist to verify that particular information has been passed along in some manner (e.g., verbally, digitally, in-person, online). 
From the interviews, it can be seen that there is a significant preference of Guideline B over Guideline A. 5 out of 6 participants chose this as the most likely to be useful for CSIRTs in varying team structures, sizes and sectors. While multiple participants expanded to then say Guideline A could be better suited to another organisational structure, these claims were apparently refuted by other participants working within those other organisational structures. Only one suggestion for Guideline A's use remains undisputed: it was proposed that it could serve well as a metric for measuring growth, particularly by management.

\subsection{Feedback on Guidelines}
It can be concluded that producing handover guidance in this manner is feasible, as there was sufficient positive feedback on the drafts, alongside general enthusiasm and assurance from participants outside the interviews. 
While specific suggestions for improvement emerged from the interviews, these were largely individual recommendations rather than consensus views, with most suggestions coming from single participants. 
We have therefore chosen not to change the guidelines based on this feedback.
This variation in feedback may reflect the diversity of organisational contexts and handover needs across different CSIRTs.
We discussed these suggestions to enable organisations that want to adopt the guidelines can integrate them into their own versions of the guidelines as required.

\subsection{Future Directions}
This exploratory study suggests several avenues for future research. 
First, a larger-scale replication would help validate and extend our findings. More significantly, we propose expanding the research into a two-phase study following the co-design framework developed by Queensland Government~\cite{MetroNorthHealth_n.d.}.

The design phase could utilise the same method as this study, using more participants. Interviews would gather feedback on the draft guidelines. It may also be useful to ask participants directly which of Patterson et al.'s (\cite{Patterson_Roth_Woods_Chow_Gomes_2004}) strategies they have encountered in their handovers. After content analysis, the feedback will be used to adjust the draft. If suggested improvements are concordant, they will be acted upon. If any suggested adjustment contradicts between participants, it will not be acted upon but instead posed in the second phase for more opinions as to whether it should be employed. 

Once these changes are made, a second design phase should ensue. The purpose of the second phase is to validate the draft created in phase 1. Participants will be asked to rate the effectiveness of the draft before and after phase 1’s development on a scale of 1--10, and give feedback on any missing or unnecessary details. They will also be given an example of a cyber incident and asked to evaluate whether the draft’s guidance would prompt a user to communicate all useful details in the scenario. The transcripts should be analysed using deductive thematic content analysis, and any adjustments should be made as previously. The final draft should then be presented in the paper, and sent to all participants.

It would also be useful to include participants from those who work \textit{with} incident responders in addition to responders themselves, in order to observe how knowledge transfers between teams.

\subsection{Limitations}
Several key limitations emerged from this study, particularly regarding sample representation and methodological constraints.
One notable takeaway from this study is the variation between organisation types, and a potential inability for those from one organisation type to speak for another. It would be beneficial to find at least one participant with experience across multiple organisation types in order to gain more understanding of how the needs of each differ, and what guidelines that satisfy all types may look like (or if this is indeed feasible). If this is not possible (or in addition), it could prove beneficial to conduct at least one group interview to allow discussion between those with experience in different settings, and identification of common ground.

We reached thematic saturation on the following interview questions: \textit{``Is your handover process optimal?''}, \textit{``Is what to include in handover taught, or do you rely upon intuition?''}, \textit{``Is there much variation in handovers between individuals?''}, \textit{``What is the main goal of a shift handover?''}, and \textit{``What difficulties are faced?''}. However, the diversity of opinion towards the handover guidelines has stopped us from doing so on the remaining questions.

The theoretical grounding of this study also presented limitations. While our findings showed alignment with Patterson et al.'s (\cite{Patterson_Roth_Woods_Chow_Gomes_2004}), strategies (strategies 1, 2 and 6 being reported as beneficial), we could have more systematically evaluated the applicability of their framework. 
The study materials could have explicitly incorporated Patterson's strategies, allowing the participants to directly assess their viability in CSIRT settings. This more structured approach to theory testing would have provided clearer insights into which strategies from other high-reliability domains transfer effectively to cybersecurity contexts.

A lot of difference between reported procedure in handovers seems to be related to the use of a ticketing system: whether a system like this is used, how it works, how tickets are filled in and how they are utilised in handovers, and so on. It would be beneficial to investigate more about these systems as they appear to be closely linked. To some degree, it appears that very similar information is conveyed in handovers between different participants; instead, the differences lie in how the information is conveyed. For example, one participant reported technical details being included in a handover, and one said this was not necessary. However, the participant who said technical details were not included then went on to say these were written in the tickets instead, which were signposted in the handover. Therefore, in some manner, the same details were conveyed in both handovers, just in different forms (direct and indirect). This suggests that research into systems like ticketing would be beneficial, as they are almost still a part of a handover, just in an indirect manner.

In addition, it was also mentioned by one participant that tasks can be accidentally handed over to an analyst that was not authorised to complete the task in some manner, meaning it had to be reassigned. This was explained to be potentially due to a lack of permissions, lack of authorised access to resources, or simply because it was not their job. This could be another element to explore: a key triad of principles in the cybersecurity domain is CIA, standing for confidentiality, integrity, and availability~\cite{Aminzade_2018}. This could therefore be another topic to investigate and how (if at all) permissions can affect handovers.

This study had a mostly varied set of participants, but was a small sample size and lacked a representation of some useful perspectives. It did not benefit from the inclusion of more than one participant with experience in an internal CSIRT, who often had opinions that disagreed with the rest of the participants. While the data collected indicates that producing guidelines is feasible, it was mentioned how much the procedures vary between organisational types and there is a potential that internal teams handover in a significantly different manner, rendering it difficult to provide guidance that applies to all. As such, it would be very beneficial to look further into the opinions of those with experience in an internal team. Therefore, if this were to be the case, adjustments could be made so that guidelines cover only external teams, for example. 

Not only this, but there was also no representation of experience with the follow-the-sun model. This between-teams handover may differ significantly from within-teams handovers; and while one participant reported experience with between-teams handovers, these were teams responsible for different jobs, rather than picking up and continuing tasks began by the previous team. This was explained in Fausett and Keebler's paper (\cite{Fausett_Keebler_2022}) as a unique difference from other domains that presents unique challenges, so ensuring guidelines cover this sufficiently (or adjusting them to explicitly cover only within-teams handovers) is another priority in future work. As such, it would be particularly beneficial to include at least one participant with experience in this.

In addition, the study lacked representation of participants working in companies with significantly flawed or failed handovers. It would be useful to include participants with this insight in the future.

In regard to methodology, this study utilised the Microsoft Teams transcription function but was not audio recorded. This led to some difficulty when the transcription software was unable to pick up particular comments, as there was no ability to go back and listen in order to resolve what was actually said. In future work, audio recording should be taken in addition to transcription through Teams or Zoom, to avoid this issue.

\section{Conclusion}
% Background
Shift handovers are critical points in which errors can arise from miscommunication. Despite this, there is little research within the cybersecurity industry. This is particularly concerning, as handovers in CSIRTs present unique challenges that can complicate matters further. \cite{Fausett_Keebler_2022} discussed the plentiful amount of research on handovers and their effectiveness in other domains, and what can be extracted and implemented in CSIRTs. For example, 80\% of errors in healthcare can be attributed to miscommunication happening specifically during handoffs~\cite{JointCommission_2012}. As mentioned, not only do other domains have significantly more research into handoffs---leading to established protocols, strategies and even standardised systems---but cyber incident response presents unique challenges that make handoffs even more complex. For example, many organisations utilise teams from different time zones to avoid night shifts, commonly referred to as the ``follow the sun'' method. This means that not only do shift handoffs occur within-teams, but also between-teams. 

% Aim 
As such, this study was designed with the aim of developing material that aids in the process of constructing handover procedure. This was an initial exploratory study, in which we worked with those in the industry to evaluate and improve guidelines produced using existing literature and resources. 
% Method
It involved interviews of six participants working in CSIRTs with experience in handovers. Transcripts of these interviews were taken, and inductive content analysis was used to investigate the data.

% Results
The analysis revealed 11 common themes. Common themes were identified as points made by at least 2 participants, and included items such as the importance of signposting, the evolution of handover procedure, individual differences in handover style and detail, and the importance of streamlining.

Guideline B was selected by 5 out of 6 participants. The guidelines were said to include all relevant detail needed, with the addition of two suggestions (a post-incident review section, and a service section for any outages or technical difficulties). However, it was reported that Guideline B needed to be condensed and made more efficient to avoid wasting time. 

% Recommendations
However, there was significant variation in opinions across multiple questions. This could be due to differences in organisation type, organisation size, or function. This highlights the benefit of future research with larger samples to reach saturation, and potentially investigate whether the differences in opinion correlate to any of these factors.

\section*{Funding and Disclosure statement}
\ifanonymous Redacted for anonymous submission \else 
Liberty Kent is supported by UK EPSRC grant no.
EP/S022503/1.
Ingolf Becker is supported by UK EPSRC grant no.
EP/W032368/1.
The authors declare that they have no known competing financial interests or personal relationships that could have appeared to influence the work reported in this paper.
\fi

\newpage
\bibliographystyle{IEEEtran}
\bibliography{bibliography}

\section{Biographies}

\textbf{Liberty Kent} is a PhD student in Cybersecurity at University College London, specialising in human factors and the performance of cyber incident response teams. She holds a degree in Experimental Psychology from the University of Oxford, and her research integrates psychology and cybersecurity to improve the effectiveness of incident response teams through cognitive support and evidence-based interventions.

\textbf{Dr Nilufer Tuptuk} is an Associate Professor in the Department of Security and Crime Science at University College London. Her research interests focus on the cybersecurity of cyber-physical systems, including critical national infrastructure systems and the study of cybercrime. Her current work involves developing AI-based models to detect and prevent cybersecurity threats and cybercrime, as well as examining vulnerabilities in AI that could be exploited to carry out malicious activities.

\textbf{Dr Ingolf Becker} is an Associate Professor in Security and Crime Science in the Department of Security and Crime Science at University College London. He is part of UCL's Academic Centre of Excellence in Cyber-Security Research (ACE-CSR) and a Principal Investigator at the Research Institute in Socio-Technical Security (RISCS). He currently lead UCL's part of the 3-year ESPRC project Protecting public-facing professionals and their dependents online (3PO).

\end{document}